\documentclass{aa}
\usepackage[varg]{txfonts}
\usepackage{graphicx}
\usepackage{ulem}
\usepackage{color}

\usepackage{natbib}
\bibpunct{(}{)}{;}{a}{}{,} 

\newcommand{\revise}[1]{#1}
\newcommand{\rev}[1]{#1}

\definecolor{orange}{rgb}{ 0.95, 0.60, 0}

\usepackage{amstext}

\begin{document}

\title{Suppression of type I migration by disk winds}
\subtitle{}
\author{Masahiro Ogihara
\and Alessandro Morbidelli
\and Tristan Guillot
}
\institute{Laboratoire Lagrange, Universit\'e C\^ote d'Azur, Observatoire de la C\^ote d'Azur, CNRS,
Blvd de l'Observatoire, CS 34229, 06304 Nice Cedex 4, France \email{omasahiro@oca.eu}
}
\date{Received 4 August 2015 / Accepted 16 October 2015}


\abstract 
{
Planets less massive than Saturn tend to rapidly migrate inward in protoplanetary disks. This is the so-called type I migration. Simulations attempting to reproduce the observed properties of exoplanets show that type I migration needs to be significantly reduced over a wide region of the disk for a long time. However, the mechanism capable of suppressing type I migration over a wide region has remained elusive. The recently found turbulence-driven disk winds offer new possibilities. 
} 
{
We investigate the effects of disk winds on the disk profile and type I migration for a range of parameters that describe the strength of disk winds. We also examine the in situ formation of close-in super-Earths in disks that evolve through disk winds.
} 
{
The disk profile, which is regulated by viscous diffusion and disk winds, was derived by solving the diffusion equation. We carried out a number of simulations and plot here migration maps that indicate the type I migration rate. We also performed \textit{N}-body simulations of the formation of close-in super-Earths from a population of planetesimals and planetary embryos.
} 
{
We define a key parameter, $K_{\rm w}$, which determines the ratio of strengths between the viscous diffusion and disk winds. For a wide range of $K_{\rm w}$, the type I migration rate is presented in migration maps. These maps show that type I migration is suppressed over the whole close-in region when the effects of disk winds are relatively strong $(K_{\rm w} \lesssim 100)$. From the results of \textit{N}-body simulations, we see that type I migration is significantly slowed down assuming $K_{\rm w}=40$. We also show that the results of \textit{N}-body simulations match statistical orbital distributions of close-in super-Earths.
} 
{}
\maketitle

\section{Introduction}
Planets with masses lower than approximately $50 M_\oplus$ (depending on the disk scale height and viscosity) migrate toward the central star under a type I migration regime. Different from linear theories,
planet formation simulations generally require the rate of type I migration to be reduced by at least a factor of ten and throughout the disk in order to reproduce the distribution of orbital distances of known exoplanets (\citealt{ida_lin08}; \citealt{ogihara_ida09}). 

It has been shown that type I migration can be locally outward depending on the disk properties (e.g., \citealt{kretke_lin12};  \citealt{bitsch_etal15}), \revise{which would change the picture of planet formation (e.g., \citealt{hellary_nelson12}; \citealt{cossou_etal14})}. The region of local outward migration arises from inhomogeneities in disks (e.g., opacity transitions). According to recent numerical simulations, it has been suggested that local inhomogeneities may help in reducing the type I migration speed (\citealt{dittkrist_etal14};\citealt{mordasini_etal15}). However, weakening of type I migration over a wide region of the disk for a long time would be required to reproduce the observed distributions of exoplanets (\citealt{ida_lin08}; \citealt{ogihara_ida09}); local traps due to disk inhomogeneities would be insufficient. 

Recent studies have shown that turbulence-driven disk winds, in which gas is blown away from the surface of the disk, can alter the density profile of the gas disk (\citealt{suzuki_inutsuka09}; \citealt{suzuki_etal10}), which can slow down or even reverse type I migration. \citet[hereafter OKIS15]{ogihara_etal15b} performed \textit{N}-body simulations of terrestrial planet formation in disks including disk winds and found that type I migration can be weakened or even reversed. They also demonstrated that characteristic features of the solar system's terrestrial planets (e.g., a mass concentration around 1 au) can be reproduced by simulations with disk winds. We anticipate that disk winds play an important role in reproducing observed orbital distributions of exoplanets by slowing type I migration over a wide range of disks for a long time.

In this work we revisit the in situ formation of close-in super-Earths in disks affected by winds. In our previous work \citep[hereafter OMG15]{ogihara_etal15a}, we reassessed the in situ formation of close-in super-Earths using \textit{N}-body simulations and observed that super-Earths undergo rapid inward migration, resulting in compact configurations near the disk inner edge, which do not match the observed distributions of super-Earths. On the other hand, we performed additional simulations in which migration is about 100 times slower. The results matched the observations much better. However, the reduction of the type I migration rate in OMG15 was just artificial and there is no physical explanation. Here we investigate whether it can be justified for disks with winds, thus providing an explanation for the observed distribution of super-Earths.

In this paper, we examine the condition for the onset of slow migration. Because of the lack of studies of disk winds, the correlation between the strength of disk winds and the resulting disk surface density slope (or type I migration rate)  has not been determined. We first investigate this by numerical experiments in Sect.~\ref{sec:migration}. Then we perform \textit{N}-body simulations of in situ formation of close-in super-Earths in a disk that evolves through disk winds in Sect.~\ref{sec:n-body}. In Sect.~\ref{sec:discussion} we give a summary.

\section{Condition for weakening of type I migration}
\label{sec:migration}

We first investigated the gas surface density slope and the type I migration rate for a wide range of parameters. We numerically solved the following diffusion equation using the same recipe as described in \citet{suzuki_etal10} and OKIS15,
\begin{equation}
\label{eq:diffusion}
\frac{\partial \Sigma_{\rm g}}{\partial t} = \frac{3}{r} \frac{\partial}{\partial r} \left[r^{1/2} \frac{\partial}{\partial r} (\nu \Sigma_{\rm g} r^{1/2})\right] - C_{\rm w} \frac{\Sigma_{\rm g}}{\sqrt{2 \pi}} \Omega,
\end{equation}
where $\Omega$ is the Keplerian frequency and  $\nu (=\alpha c_{\rm s} H)$ is the viscosity. We used the $\alpha$-prescription for the viscosity, where $c_{\rm s}$ and $H$ indicate the sound velocity and the disk scale height, respectively. The disk wind flux ($\rho v_z$) can be expressed as $C_{\rm w} \rho_0 c_{\rm s} = C_{\rm w} \Sigma_{\rm g} \Omega /\sqrt{2 \pi}$ using the mid-plane density $\rho_0$ and a non-dimensional constant $C_{\rm w}$ \citep{suzuki_etal10}. The initial condition for the gas disk is $\Sigma_{\rm g} = 2400 (r/1 \rm{au})^{-3/2} \exp(-r/50 {\rm au})\,\mathrm{g\, cm}^{-2}$. The temperature profile is assumed to be that of \citet{hayashi81} as $T = 280 (r/1 {\rm au})^{-1/2} {\rm K}$.

\begin{figure}
\resizebox{0.9 \hsize}{!}{\includegraphics{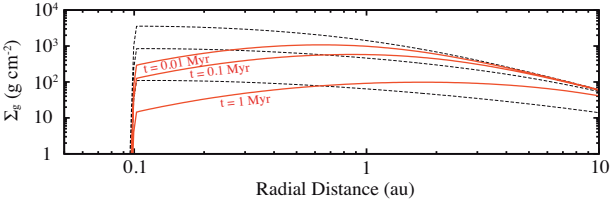}}
\caption{Evolution of gas surface density profile for $t = 0.01 {\rm Myr}, 0.1 {\rm Myr},$ and $1 {\rm Myr}$. Dotted lines show the case of weak disk winds $(K_{\rm w}=200)$. Solid lines indicate the case of strong disk winds $(K_{\rm w}=40)$.}
\label{fig:r-sigma}
\end{figure}

Figure~\ref{fig:r-sigma} shows examples of the gas surface density evolution. To highlight the difference between the simulation including disk winds and those in OMG15 in the next section, a disk inner edge at $r = 0.1 {\rm au}$ is superposed to the gas surface density. It is readily seen that the disk profile is altered, especially in the close-in region. The dotted lines indicate the disk evolution for $\alpha = 10^{-3}$ and $C_{\rm w}=5\times10^{-6}$. The surface density slope is gentle inside $r=1 {\rm au}$ and almost flat at $r = 0.1 {\rm au}$. Solid lines represent the evolution for $\alpha = 10^{-4}$ and $C_{\rm w}=2.5\times10^{-6}$. The slope of the surface density of the gas is positive inside $r=1{\rm au}$. 

We here introduce a parameter $K_{\rm w} (\equiv \alpha/C_{\rm w})$ for later discussion. The mass transport rate due to viscous transport and the mass-loss rate due to disk winds in an annulus with $\Delta r$ are
\begin{eqnarray}\Delta \dot{M}_{\rm vis} = \frac{\partial}{\partial r}\left[-3 \pi \left(\Sigma_{\rm g} \nu + 2 r \frac{\partial \Sigma_{\rm g}\nu}{\partial r} \right)\right] \Delta r,\\
\Delta \dot{M}_{\rm wind} = -2 \pi r C_{\rm w} \frac{\Sigma_{\rm g}}{\sqrt{2 \pi}} \Omega \Delta r,
\end{eqnarray}
respectively. When $\Delta \dot{M}_{\rm vis} > \Delta \dot{M}_{\rm wind}$ the disk evolution is dominated by the viscous transport. Here,
\begin{eqnarray}
\frac{\Delta \dot{M}_{\rm vis}}{\Delta \dot{M}_{\rm wind}} \simeq \frac{9 \sqrt{\pi}}{2} \left(\frac{H}{r}\right)^2 \frac{\alpha}{C_{\rm w}}
\simeq 0.02 \left(\frac{r}{1 {\rm au}} \right)^{1/2} K_{\rm w},
\end{eqnarray}
meaning that disk winds become significant inside 1\,au when $K_{\rm w} \lesssim 100$. Disk evolution of dashed lines and solid lines in Fig.~\ref{fig:r-sigma} correspond to $K_{\rm w}=200$ and 40, respectively. 

Next, by performing a number of simulations for a wide range of parameters, we determined the efficiency of type I migration. The surface density slope was determined by viscous diffusion and mass loss due to disk winds; a key parameter is $K_{\rm w}$. The total torque for type I migration is given by
\begin{eqnarray}
\label{eq:torque}
\Gamma =  \frac{\beta}{2}  \left(\frac{M}{M_*}\right)
\left(\frac{\Sigma_{\rm g} r^2}{M_*}\right)
\left(\frac{c_{\rm s}}{v_{\rm K}}\right)^{-2} M v_{\rm K}^2,
\end{eqnarray}
where $\beta, M, M_*,$ and $v_{\rm K}$ are a coefficient that determines the direction and rate of type I migration, the mass of the planet, the mass of the host star, and the Keplerian velocity, respectively. For details of expression of $\beta$, we refer to Eqs.~(11)-(13) in OKIS15, which is based on \citet{paardekooper_etal11}. When the surface density slope and the temperature gradient are given, $\beta$ is determined by the saturation of corotation torque. The level of saturation is expressed by the parameter
\begin{eqnarray}
\label{eq:pnu}
P_\nu = \frac{2}{3} \sqrt{\frac{\Omega r^2 x_s^3}{2 \pi \nu}},
x_s =  \frac{1.1}{\gamma^{1/4}} \sqrt{\frac{M}{M_*} \frac{r}{H}},
\end{eqnarray}
where $x_s$ is the dimensionless half-width of the horseshoe region and $\gamma$ is the adiabatic index. When the temperature distribution is fixed, $P_\nu$ is determined by viscosity and planetary mass. The saturation of the entropy-related corotation torque is determined by the thermal diffusivity $\xi$, and we assumed $\xi = \nu$ for simplicity. 

\begin{figure}
\resizebox{0.95 \hsize}{!}{\includegraphics{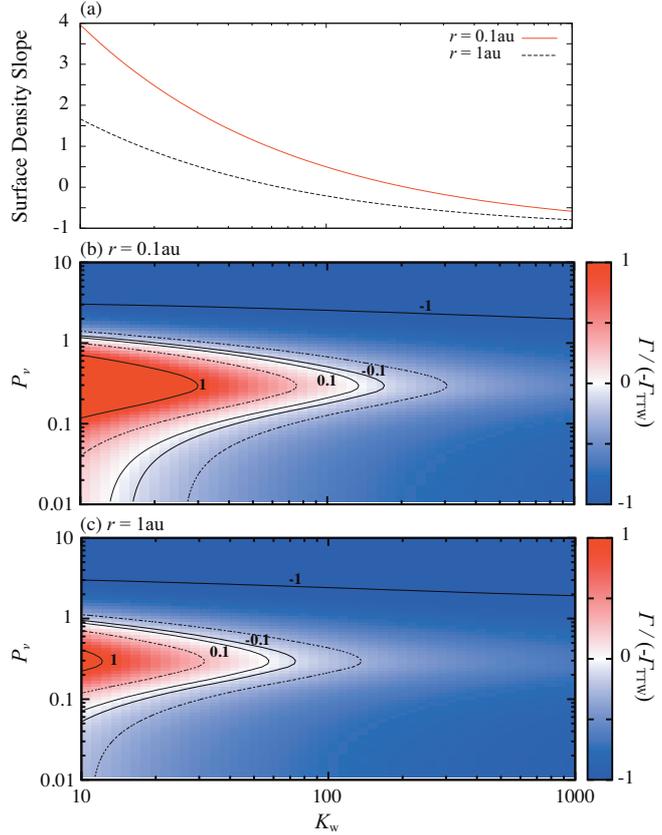}}
\caption{Surface density slope in a steady state at $r=0.1 {\rm au}$ and $1 {\rm au}$ for various values of $K_{\rm w}$ (panel (a)). Migration efficiency $(\Gamma / (-\Gamma_{\rm TTW}))$ for planets with $e=0.01$ at $r=0.1 {\rm au}$ (panel (b)) and $r=1{\rm au}$ (panel (c)). The contours show a migration efficiency of -1, -0.3, -0.1, 0.1, 0.3, and 1. When $\revise{\Gamma > 0}$, planets move outward.}
\label{fig:map}
\end{figure}

Figure~\ref{fig:map}(a) represents the surface density slope at $r=0.1 {\rm au}$ (the surface density slope just beyond the disk edge at 0.1\,au) and $1 {\rm au}$. Figure~\ref{fig:map}(b) and (c) shows migration maps of $r=0.1 {\rm au}$ and $1 {\rm au}$, respectively. The color scale indicates the migration efficiency as compared to the migration rate in a locally isothermal disk with a power-law index of -3/2 derived by a three-dimensional linear analysis by \citet[hereafter TTW02]{tanaka_etal02}, which is defined by $\Gamma/(-\Gamma_{\rm TTW})$\footnote{By using a commonly used value of $\Gamma_0 = (M/M_*)^2 (c_{\rm s}/v_{\rm K})^{-2} \Sigma_{\rm g} r^2 v_{\rm K}^2$, the efficiency is also expressed by $\Gamma/(-\Gamma_{\rm TTW}) = \Gamma/(2.175\Gamma_0)$.}. $\Gamma_{\rm TTW}$ has a negative value, so $\Gamma >0$ indicates outward migration. This value corresponds to the negative of the efficiency parameter, $-C_{\rm 1}$, used in \citet{ida_lin08}. We included the dependence of the corotation torque on the eccentricity by assuming $e=0.01$ \citep{fendyke_nelson14}. We adopted a higher value for $C_{\rm w}(=10^{-4})$ in Fig.~\ref{fig:map} to reduce the computation time. As already stated above, the surface density slope is determined only by $K_{\rm w}$, which is confirmed by numerical experiments that adopt different values for $C_{\rm w}$. The surface density slope relaxes to a steady state after $t=t_\nu = r^2/\nu$, where $t_\nu$ is the viscous timescale, so the maps are plotted at $t=10^5 {\rm yr}$.

From Fig.~\ref{fig:map}(a), we find that the surface density slope decreases as $K_{\rm w}$ increases. Thus, a smaller $K_{\rm w}$ yields slower or even outward migration. At $r=0.1 {\rm au}$ for $K_{\rm w} \lesssim 150$, we can find a range of $P_\nu$
in
which the migration speed is reduced by a factor of more than ten relative to that predicted by TTW02. As $r$ increases, the deviation of the slope from that of initial power-law disks decreases. At $r=1 {\rm au}$, the migration can be reduced by a factor of ten from TTW02 for $K_{\rm w} \lesssim 70$. Thus, disk winds are able to modify type I migration in a wide region inside 1\,au.

The migration rate depends not only on $K_{\rm w}$ but also on $P_\nu$. The value of $P_\nu$ increases with decreasing viscosity; the corotation torque saturates at low viscosity, and the Lindblad torque dominates the total torque. On the other hand, it is also known that there exists a cut-off for the horseshoe drag at high viscosity (small $P_\nu$) and the corotation torque approaches its linear value (e.g., \citealt{masset02}; \citealt{paardekooper_papaloizou09a}).

Using Fig.~\ref{fig:map}, we can estimate the migration rate for different sets of parameters ($K_{\rm w}, P_\nu$). According to \citet{suzuki_etal10}, a possible value of $K_{\rm w}$ would be $\sim 100$; however, further investigation would be required by global magnetohydrodynamics simulations that cover \revise{enough grid points in the vertical direction} to constrain the range. 

Our results depend strongly on the fact that $K_{\rm w}$ is constant with time and radius. As discussed by \citet{suzuki_etal10}, the gravitational energy is released by gas accretion, and a part of the energy is used to drive winds. Thus $C_{\rm w}$ is proportional to the kinetic energy of winds, while $\alpha$ is proportional to the accretion rate. Thus $K_{\rm w} = {\rm const}$ is a natural assumption\footnote{
A non-uniform $K_{\rm w}$ may arise in the case of strong disk winds, for example, which would cause an inner cavity. A dead zone in the disk would likewise lead to non-uniformity (\citealt{suzuki_etal10}). These two cases are not considered here, but we note that the latter gives rise to a density profile that is very similar to the case without a dead zone. 
}. We note that although we assumed that gas blown out of the disk surface escapes from the disk, some materials may return to the disk. Stellar winds can push away the lifted-up gas \citep{suzuki_etal10}; however, if a large amount of gas returns to the disk, the surface density slope would be smaller than in Fig.~\ref{fig:map}(a). Moreover, we assumed a weak vertical magnetic field ($\beta_z \gtrsim 10^4$; $\beta_z$ is the vertical component of plasma $\beta$ at the midplane).

\revise{As discussed above, our model uses the same thermal profile for the disk as in OMG15. Clearly, this is a simplification that we adopted to compare the results to those of OMG15 more directly. However, we checked that our main results would hold with a more realistic temperature profile (\citealt{bitsch_etal15}, B. Bitsch, private communication). Specifically, we found that while the temperature and scale height are lower than for our model for $\alpha = 10^{-4}$, the changes in the migration maps remain limited compared to those shown in Fig.~\ref{fig:map}. In particular, we confirm that type I migration is still suppressed for some value of $P_\nu$ for $K_{\rm w}<100$. Furthermore, if a density gap is opened by planets, the formulae we used would not be valid. While in our disk model only massive super-Earths $(\gtrsim 20 M_\oplus)$ would open a gap, in the disk with a more realistic temperature profile (i.e., colder), even lower-mass planets $(\gtrsim 2 M_\oplus)$ open a gap. However, in this case the planets would migrate in the type II regime, which, for the value of $\alpha = 10^{-4}$ that we assumed, would result in a slower migration rate than in our disk model. Thus the results we present in the next section concerning the effects of a reduced migration speed are conservative in the sense that the migration speed of the most massive planets would be even lower in a disk with a more realistic temperature model.}

\section{In situ formation of close-in super-Earths}
\label{sec:n-body}
We now perform ed\textit{N}-body simulations of formation of close-in super-Earths in disks that evolve through disk winds. The simulation model is the same as that of OMG15, except for the gas disk model. According to a power-law distribution, 250 embryos with a mass of $0.1 M_\oplus$ and 1250 planetesimals with a mass of $0.02 M_\oplus$ were distributed between 0.1 and 1\,au. The total mass in the system was set to $50 M_\oplus$. \rev{This high solid-to-gas mass ratio can be explained by invoking the radial drift of dust and pebbles into the inner part of the disk, but we place ourselves here at a stage where most of the dust has already been converted into large bodies (embryos and planetesimals).} Planetesimals suffer aerodynamical gas drag \revise{assuming a physical size of 50 km in radius} \citep{adachi_etal76}, while planets with more than roughly $0.1 M_\oplus$ undergo the tidal damping of eccentricities, inclinations, and semimajor axes (see \citealt{ogihara_etal14} and OKIS15 for each formula).

To show the effects of disk winds in a disk model, we assumed that disk winds are relatively strong ($K_{\rm w}=40$), where $\alpha = 10^{-4}$ and $C_{\rm w} = 2.5 \times 10^{-6}$ were used. The evolution of the gas surface density is shown by solid lines in Fig.~\ref{fig:r-sigma}. In this disk, we see from Fig.~\ref{fig:map} that type I migration can be slower by a factor of more than ten from TTW02 or even reversed for $0.1\lesssim P_\nu\lesssim1$. This range roughly corresponds to $M \sim 0.1-1 M_\oplus$ between $r = 0.1$ and $1 {\rm au}$. We assumed a rapid disk dispersal ($\sim 0.1 {\rm Myr}$) after a typical disk lifetime of $3 {\rm Myr}$ to be consistent with observations.

\begin{figure}
\resizebox{0.9 \hsize}{!}{\includegraphics{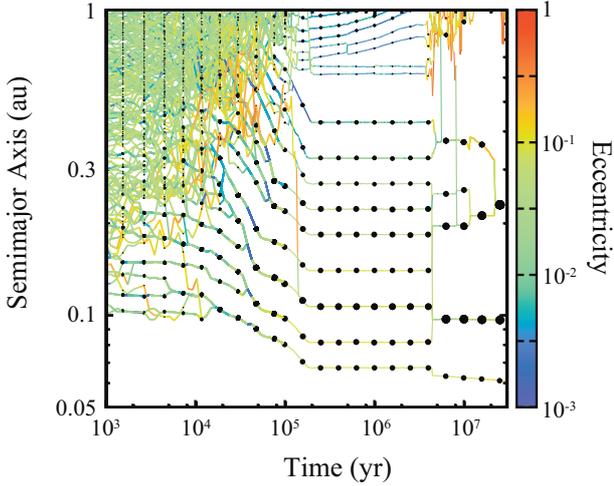}}
\caption{Time evolution of planets for a typical run.
The filled circles connected with solid lines represent the sizes of planets. The smallest circle represents an embryo of 0.2 Earth-mass, while the largest circle represents a 20 Earth-mass planet. The color of the lines indicates the eccentricity (color bar).}
\label{fig:t-a}
\end{figure}

We performed ten runs with different initial positions of solid bodies; a typical run is shown in Fig.~\ref{fig:t-a}. As seen in OMG15, the growth of embryos is quite rapid. However, we find that the migration speed is significantly slower than in the fiducial model (model~1) in OMG15 (see Fig.~2 in OMG15). The actual migration timescale of massive planets $(\sim 5 M_\oplus)$ is $\gtrsim 0.1 {\rm Myr}$ between $t = 0.01$ and 0.1 Myr. This migration rate is about a few to ten times slower than that predicted by TTW02. At $t \simeq 0.2 {\rm Myr}$, planets are captured in mutual mean motion resonances, making a long resonance chain (nine bodies). The chain undergoes orbital instability at about 4 Myr after gas dispersal, leading to mutual collisions and a relatively separated system of three planets out of resonances. Compared with the results of slow-migration case (model~3) in OMG15, the migration speed is faster, and hence there are fewer planets in the resonant chain in our results. However, the final phase of the planet formation process, in which planets undergo close encounters and collisions after gas dispersal, is quite similar.

\begin{figure}
\resizebox{0.8 \hsize}{!}{\includegraphics{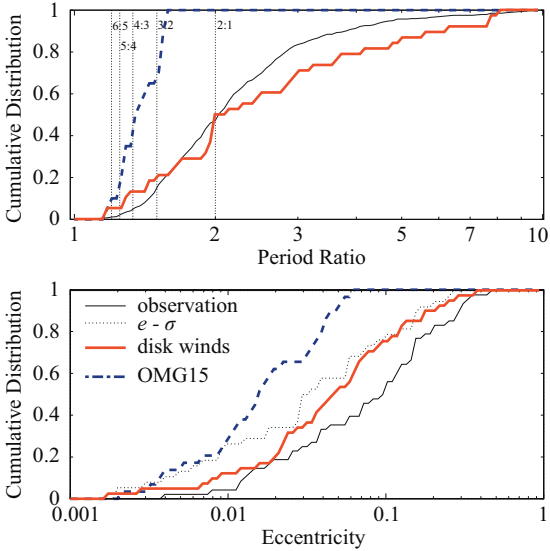}}
\caption{Comparison of cumulative period ratio distributions and eccentricity distributions. Thin solid lines show observed distributions of confirmed close-in super-Earths as of June 2015 (341 systems with 854 planets). Thick solid lines indicate results of simulations including the effects of disk winds, while thick dashed lines show the previous results for model~1 in OMG15. As discussed in OMG15, observed eccentricities can be overestimated (e.g., \citealt{shen_turner08}; \citealt{zakamska_etal11}). The thin dotted line shows the eccentricity distribution, in which each eccentricity is assumed to be $e - \sigma$. \revise{Here, $\sigma$ is the estimated error}.
}
\label{fig:p-n}
\end{figure}

Figure~\ref{fig:p-n} compares the results of simulations with the observed distributions of period ratios of adjacent pairs and eccentricities. For reference, the results of a fiducial model (model~1) in OMG15 are also plotted. We find that the period ratio distribution of simulations including disk winds matches the observations much better than that of model~1 in OMG15. The eccentricity distribution also matches the observations well. We acknowledge, however, that the observed mass distribution is very poorly matched by our simulations; the averaged slope of solid surface density in our simulation, which is deduced from the distribution of the final planets, is $\sim -3$, which is significantly steeper than the averaged slope of close-in super-Earths ($\simeq -1.5$). 
This is presumably because we assumed an initial distribution of solids that is confined to a region between 0.1 and 1\,au, so the slope would be improved if planets with $M \gtrsim M_\oplus$ (in this case $\Gamma < 0$ at 1\,au) migrated from outside 1\,au.

Here we briefly discuss a successful scenario of in situ formation of close-in super-Earths. According to the results of \textit{N}-body simulations presented in this and previous papers, in a successful model planets are captured in a resonance chain in a disk and then undergo orbit crossings and collisions during disk dissipation. On the other hand, unsuccessful models include the rapid migration case (e.g., model~1 in OMG15) and the no migration case (e.g., model~4 in OMG15). In the former case, planets formed in a compact resonance chain with a small number of planets ($N \simeq 5$). The small number of planets in the chain prevents close encounters after gas dispersal, leading to mismatches in the period ratio (too compact) and eccentricity (too low). In the latter case, planets undergo a too violent instability because they are not in a resonant chain, which also results in mismatches to observations (too separated and high eccentricities).

To realize the successful scenario described above, the number of planets in a resonance chain should be large enough to trigger orbit crossings during disk dissipation. According to a study on the orbital stability of a resonance chain \citep{matsumoto_etal12}, there should be more than five to ten planets in a chain. It is difficult to precisely assess the conditions for the formation of a resonant chain with $N > 5-10$; however, results of \textit{N}-body simulations imply that type I migration should be reduced by a factor of about ten from that predicted by the linear theory. According to Fig.~\ref{fig:map}, this condition corresponds to $K_{\rm w} \lesssim 100$ and $0.1 \lesssim P_\nu \lesssim 1$.

\section{Summary}
\label{sec:discussion}

We computed the surface density profile of disks affected by winds of various strengths and the resulting type I migration rates. We confirmed that the type I migration can be slowed down in the whole close-in region $(r < 1 {\rm au})$ if the wind is sufficiently strong. Using the migration map in Fig.~\ref{fig:map}., the migration rate for different sets of parameters can be estimated without the need for additional calculations. We also performed \textit{N}-body simulations of the formation of close-in super-Earths that included the effects of disk winds. Given that the effects of disk winds are relatively strong, we demonstrated that type I migration is significantly slowed down. This is the first simulation in which the observed statistical orbital distributions of close-in super-Earths are reproduced by results of \textit{N}-body simulations without applying an artificial reduction of type I migration.

\begin{acknowledgements}
We thank the anonymous referee for helpful comments and T. Suzuki for valuable discussions. This work was supported by ANR, project number ANR-13--13-BS05-0003-01 projet MOJO (Modeling the Origin of JOvian planets).
\end{acknowledgements}

{}






\end{document}